\documentclass[journal]{IEEEtran}
\ifCLASSINFOpdf
\else
\fi
\usepackage{mathrsfs}
\usepackage{graphicx}
\usepackage{epsfig}
\usepackage{amsmath}
\usepackage{bm}
\usepackage{float}
\usepackage{multirow}
\usepackage{color}
\usepackage{subfig}
\usepackage{amsmath}
\usepackage{algorithm}
\usepackage{algorithmic}
\usepackage{cite}
\usepackage{amsthm}
\usepackage{amssymb}
\usepackage{tabularx}
\usepackage{booktabs}

\newenvironment{sequation*}{\begin{equation*}\small}{\end{equation*}}

\theoremstyle{definition}

\theoremstyle{remark}

\theoremstyle{proposition}

\long\def\symbolfootnote[#1]#2{\begingroup%
\def\thefootnote{\fnsymbol{footnote}}\footnote[#1]{#2}\endgroup}

\ifCLASSOPTIONcompsoc
\usepackage[nocompress]{cite}
\else
\usepackage{cite}
\fi

\begin{document}

\bibliographystyle{IEEEtran}

\title{Dynamic Resource Allocation in Next Generation Cellular Networks with Full-Duplex Self-backhauls}
\author{\normalsize \IEEEauthorblockN{Lei Chen\IEEEauthorrefmark{1}, F. Richard Yu\IEEEauthorrefmark{1}, Hong Ji\IEEEauthorrefmark{2}, Bo Rong\IEEEauthorrefmark{3}, and Victor C.M. Leung\IEEEauthorrefmark{4}}\\
\IEEEauthorrefmark{1}Depart. of Systems and Computer Eng., Carleton Univ., Ottawa, ON, Canada\\
\IEEEauthorrefmark{2}Key Lab. of Universal Wireless Comm., Beijing Univ. of Posts and Telecom., P.R. China\\
\IEEEauthorrefmark{3}Communications Research Centre, Ottawa, ON, Canada\\
\IEEEauthorrefmark{4}Depart. of Electrical and Computer Eng., The Univ. of British Columbia, Vancouver, BC, Canada\\
\thanks{This paper is jointly supported by the Hi-Tech Research and Development
Program of China (National 863 Program) under Grant 2014AA01A701 and National Natural Science Foundation of China under Grant 61271182.}}

\maketitle

\begin{abstract}
With the dense deployment of small cell networks, low-cost backhaul schemes for small cell base stations (SBSs) have attracted great attentions. Self-backhaul using cellular communication technology is considered as a promising solution. Although some excellent works have been done on self-backhaul in small cell networks, most of them do not consider the recent advances of full-duplex (FD) and massive multiple-input and multiple-output (MIMO) technologies. In this paper, we propose a self-backhaul scheme for small cell networks by combining FD and massive MIMO technologies.  In our proposed scheme, the macro base station (MBS) is equipped with massive MIMO antennas, and the SBSs have the FD communication ability. By treating the SBSs as \textit{special} macro users, we can achieve the simultaneous transmissions of the access link of users and the backhaul link of SBSs  in the same frequency. Furthermore, considering the existence of inter-tier and intra-tier interference, we formulate the power allocation problem of the MBS and SBSs as an optimization problem. Because the formulated power allocation problem is a non-convex problem, we transform the original problem into a difference of convex program (DCP) by successive convex approximation method (SCAM) and variable transformation, and then solve it using a constrained concave convex procedure (CCCP) based iterative algorithm. Finally, extensive simulations are conducted with different system configurations to verify the effectiveness of the proposed scheme.
\end{abstract}


\section{Introduction}

With the explosive demand for mobile broadband services and the emergence of new high capacity mobile devices, mobile networks have to continuously evolve to meet capacity and coverage demands with the latest technologies. At the same time, there is an apparent trend of declining profitability of mobile data despite the recent exponential growth of mobile data usage \cite{T11A}. Small cell networks, which can improve spectrum efficiency, energy efficiency, and coverage effectively, as well as reduce the capital expenses (CapEx) and operation expenses (OpEx), have been considered as an important technology of next generation cellular networks \cite{BY14,F15M,LY15}. Furthermore, dense small cell networks have attracted great attentions, where a mass of small cell base stations (SBSs) are deployed to improve the quality of service (QoS) further \cite{J14C}. However, with the dense deployment of massive SBSs, the \emph{backhaul} problem becomes severe increasingly. Therefore, optimizing backhaul is critical to satisfy the QoS of users and reduce the CapEx and OpEx.

The backhaul technologies for small cells can be classified into three categories. The first is the wired optical fibre, which has high capacity and will undoubtedly connect a major portion of small cells, especially in the long run \cite{R13D}. The second is wireless point-to-point microwave \cite{S13M} or mmWave \cite{T15P}, which uses high-gain directional antennas in line-of-sight (LOS) environments and provides high-capacity backhaul link. Unfortunately, installing fibres, microwave or mmWave equipment is expensive and time-consuming, preventing fast deployment of SBSs \cite{Mahloo2014Cost}. In addition, the microwave and mmWave usually operate in the LoS environment and are not suitable for the urban environment because of the massive buildings. The third possible backhaul technology is the cellular communication technology (e.g., LTE). It uses the cellular spectrum to access and backhaul, and is suitable for the non-LoS environment because of the radio nature of cellular spectrum \cite{B13S}. Moreover, with the cellular communication technology backhaul, which is called \textit{self-backhaul} in the literature (e.g., \cite{SSGBRW13}), the SBSs do not require extra backhaul equipment or spectrum, and consequently, self-backhaul is a promising technology in future small cell networks \cite{SSGBRW13}.

Nevertheless, due to the limited cellular spectrum and the existence of inter-tier and intra-tier interference, designing self-backhaul schemes is challenging. In \cite{H13J}, the authors proposed a multi-hop self-backhaul scheme by jointly considering resource allocation and routing. The authors of \cite{L06F} studied the fair scheduling problem in wireless multi-hop self-backhaul networks. The synchronization issue of time division duplex (TDD)-based self-backhaul was studied in \cite{B13S}. Although those works are excellent, they focus on the traditional half-duplex self-backhaul. With the development of self-interference cancellation technologies, full-duplex (FD) communication technology becomes possible. FD makes it possible for radios to transmit and receive simultaneously in the same frequency band, which nearly doubles the spectrum efficiency\cite{LiuGang2014}. {The FD relay has been researched in many papers \cite{H2014Low}\cite{I2012Full} and the} authors of \cite{R2015Full} first proposed FD self-backhaul scheme where FD communication hardware is equipped in SBSs. Consequently, SBSs receive data from the macro base station (MBS) and transmit it to SBSs in downlink (DL); at the same time, they receive data from mobile users in uplink (UL) and transmit it to the MBS and  in same frequency. Moreover, the effectiveness of FD self-backhaul scheme had been proved in this paper. However, the authors do not consider the interference problem in small cell networks. In particular, massive MIMO technology is not involved in this paper.


 Massive MIMO technology can achieve the transmission of multiple users at the same time and in the same frequency band by utilizing the large spatial degrees-of-freedom, which will also improve the spectrum efficiency of the system \cite{L14A}. Due to the fact that the MBS is responsible for the transmission to SBSs besides its users in FD self-backhaul scheme, the MBS with massive MIMO can transmit or receive data to SBSs and its users simultaneously in the same frequency band. 
In this way, the spectrum efficiency of backhaul and access links will be improved significantly.
In \cite{H2013Massive},\cite{H2013Making}, the authors intended to analyze the gain of jointly considering massive MIMO and small cells, but they did not consider the backhaul problem of small cell networks and simplified the power allocation of MBS and SBSs by equal power allocation. 

{Despite the potential vision of FD self-backhaul small cell networks with massive MIMO, many research challenges remain to be addressed. One of the main research challenges is resource allocation, which plays an important role in traditional wireless networks \cite{Laks2015Transmit,YWL06_MONET,LYJ10,MYL04,YL01,LYH10,YK07,XYJL12,BYC12,XYJ12}. When FD self-backhaul and massive MIMO are jointly considered, the problem of resource allocation becomes even more challenging. On one hand, the FD self-backhaul makes the backhaul and access links coupled, which depends on power allocation and self-interference cancellation performance. On the other hand, massive MIMO will introduce the interference among multiple users. To the best of our knowledge, the problem of power allocation in FD self-backhaul small cell networks with massive MIMO has not been studied in previous works. The distinct features of this paper are summarized as
follows}

\begin{itemize}
\item We propose a novel architecture of FD self-backhaul small cell networks with massive MIMO technology. In our proposed scheme, the MBS is equipped with massive MIMO antennas, and the SBSs have the FD capability. By treating the SBSs as \textit{special} macro users, we can achieve the transmissions of the access link of users and the backhaul link of SBSs simultaneously in the same frequency band.
\item Furthermore, considering the existence of inter-tier and intra-tier interference, we formulate the power allocation problem of the MBS and SBSs as an optimization problem, which maximizes the total spectrum efficiency (SE) of the small cell networks, while considering the lowest quality of service (QoS) requirement of users. In addition, we take the residual self-interference of FD communications into account in the formulated problem.
\item Since the formulated problem is a non-convex optimization problem, its computational complexity is high. To solve it efficiently, we {transform} the original problem into a difference of convex program (DCP) by using successive convex approximation method (SCAM) and appropriate variable substitution, and then solve it using a constrained concave convex procedure (CCCP)-based iterative algorithm, which reduces the computation complexity significantly. Furthermore, we prove the convergence of our proposed iteration algorithm.
\item Extensive simulations are conducted with different system configurations to verify the effectiveness of the proposed FD self-backhaul scheme with massive MIMO. 
\end{itemize}



The rest of this paper is organized as follows. The proposed FD self-backhaul scheme with massive MIMO is described in Secion \ref{Network model}. The power allocation problem is formulated in Section \ref{Problem formulation}. Then we solve the optimization problem in Section \ref{Solution of the optimization problem}. Simulation results are discussed in Section \ref{Simulation}. Finally, we conclude this study and describe our future works in Section \ref{Conclusion}.

\section{Network model}
\label{Network model}
\begin{figure}[!t]
\centering
\includegraphics[width=0.5\textwidth]{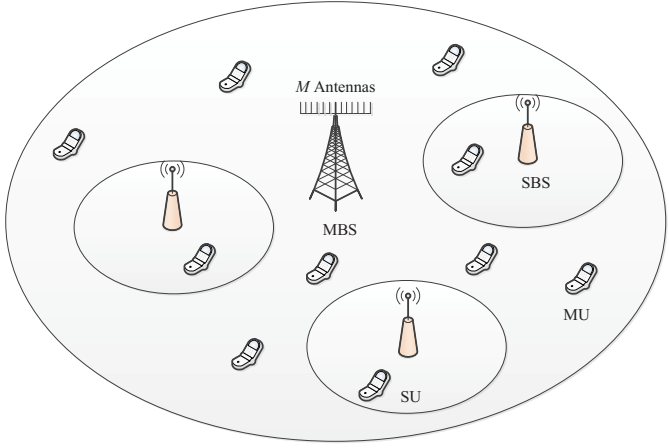}
\caption{Network framework.}
\label{fig: Network model}
\end{figure}

In this paper, we consider a two-tier small cell network consisting of one MBS with $M$ antennas and $N$ SBSs with single antenna\footnote{To simplify the network model, single-antenna SBS is assumed in this paper and the works of this paper can be expanded to small cell networks with multi-antenna SBS by involving multi-antenna channel model.}, as shown in Fig. \ref{fig: Network model}. To fully exploit the spectrum resource, the MBS and SBSs share the spectrum. For ease of presentation, we assume that the MBS serves $K$ single-antenna users (MUs) and each SBS serves one single-antenna user (SU). Note that {$K+N\ll M$}, which means that a massive MIMO system is adopted in this paper and $M$ could be very large (e.g., 100, 1000, or even more \cite{R2013Scaling}).
We define by the set $\mathcal{U}^m$ and the set $\mathcal{U}^s$ the MUs and SUs, respectively, where the element $u_k^m$ of $\mathcal{U}^m$ and $u_n^s$ of $\mathcal{U}^s$ represent the $k$-th MU and the SU of the $n$-th small cell BS, respectively. To reap the benefits of massive MIMO antennas, channel state information (CSI) must be available at transmitter, the TDD protocol is adopted in this paper because the channel reciprocity can be exploited\footnote{The proposed FD self-backhaul scheme with massive MIMO also can be used in FDD system based on some channel estimation method \cite{Zhi2015Achievable}, \cite{Zhen2015Block}.}, which allows the MBS to estimate its DL channel from UL pilots sent by the users. In this paper, {we assume that all base stations (BSs)} serve users over flat-fading channel.
\subsection{FD Self-backhaul Scheme of Small Cell Networks with Massive MIMO}
\begin{figure}[!t]
\centering
\includegraphics[width=0.5\textwidth]{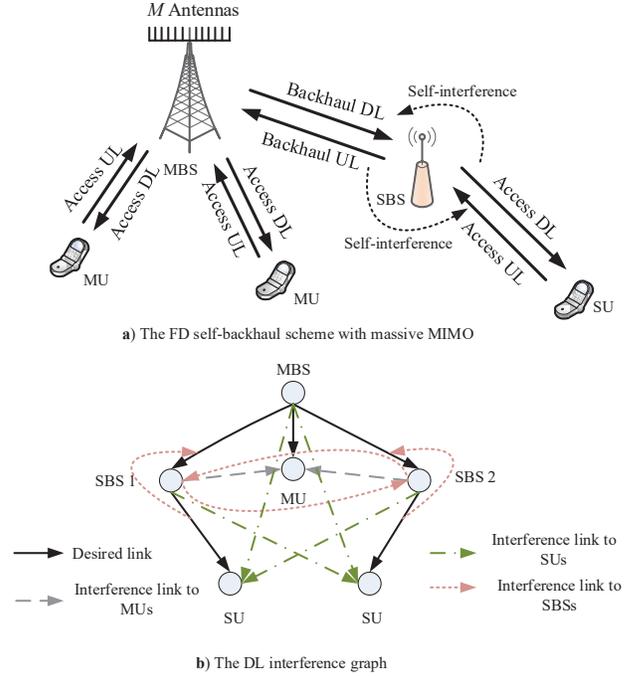}
\caption{The FD self-backhaul scheme with massive MIMO and the DL interference graph.}
\label{fig: self-bachhaul and interference graph}
\end{figure}

As shown in Fig. \ref{fig: self-bachhaul and interference graph}(a), SBSs are equipped with FD hardware, which enables them to backhaul data for themselves. In the DL, a SBS can receive data from the MBS while simultaneously transmitting to its users at the same resource block. In the UL, a SBS can receive data from the users while simultaneously transmitting data to the MBS at the same resource block. Note that the In this mechanism, the small cell can effectively backhaul itself, eliminating the need for a separate backhaul solution or a separate backhaul frequency band. Therefore, self-backhauling can significantly reduce the cost and complexity of rolling out small cell networks. In order to distinguish DL from UL in access and backhaul transmissions, we call the relevant links as access UL, access DL, backhaul UL, and backhaul DL, respectively. Due to the limitation of self-interference cancellation technologies, the backhaul DL and access UL will suffer some self-interference from access DL and backhaul UL, respectively.

In massive MIMO systems, the BS can transmit or receive data to/from multiple users simultaneously in the same frequency band by using the beamforming technology or by using virtual MIMO technology, respectively. We study the FD self-backhaul in  small cell networks with massive MIMO, where the SBSs are considered as \textit{special MUs}. In the DL, the MBS transmits data to MUs and SBSs, and SBSs transfer the data to their users simultaneously in the same frequency band. In the UL, the SBSs receive data from their users and transfer them to the MBS, and the MBS receives data from MUs and SBSs simultaneously in the same frequency band. In other words, by jointly considering FD and massive MIMO technologies, we achieve not only the transmission or reception of MUs and SUs in the same frequency band at the same time, but also the access and backhaul of SBSs in the same frequency band at the same time. This scheme will improve the spectrum efficiency and decrease the cost of backhaul infrastructure. Comparing with DL, UL usually has less traffic, so we focus on the transmission of DL in this paper. For ease of analysis, full buffer traffic model is assumed in the MBS and SBSs. Note that pilot contamination problem in massive MIMO system is not considered due to the fact that many excellent works have been done.
\subsection{Channel Model}

The channel matrix from the transmission node to reception node can be written as $\mathbf{G}=\mathbf{D}^{1/2}\mathbf{H}$, where $\mathbf{D}=diag\{\beta_1, \beta_2, ...., \beta_I\}$ ($I$ indicates the number of reception nodes, $I=1$ for SBSs' DL and $I=K+N$ for MBS' DL). The component $\beta_i=\varphi\zeta/d_i^{\alpha}$ consists of path loss and shadow fading, $\varphi$ is a constant related to carrier frequency and antenna gain, $d_i$ is the distance between the two nodes, $\alpha$ is the path loss exponent, $\zeta$ represents the shadow fading which follows the log-normal distribution $10\log\zeta\sim N(0, \sigma^2)$. Fast fading matrix is $\mathbf{H}=[\mathbf{h}_1^T, \mathbf{h}_2^T, ..., \mathbf{h}_I^T]^T\in C^{I\times J}$ ($J$ represents the number of antennas of transmission node, $J=1$ for SBSs' DL and $J=M$ for MBS' DL), the components $h\sim N(0, 1)$ are Rayleigh flat-fading random variables. In this paper, the Zero-Forcing Beamforming is adopted for MBS DL. With it, the multi-user interference can be eliminated perfectly \cite{T06On}. To reduce the complexity and make beamforming scheme effective, the Zero-Forcing Beamforming only be operated among MUs' DLs without including the backhaul DLs. $\mathbf{W}=[\mathbf{w}_1, \mathbf{w}_2, ..., \mathbf{w}_{K+N}]\in C^{M*(K+N)}$ is defined as the precoding matrix of the MBS. For easy reading, the notations of this paper are described as in Table \ref{table_1}.
\begin{table}[!t]
\centering
\caption{Notation}
\label{table_1}
\begin{tabular}{ll}
\toprule
Notation & Definition\\
\midrule
$K$                 & The number of MUs            \\
${M}$               & The number of the MBS' antennas  \\
$N$                 & The number of SBSs            \\
$P$                 & Transmission power        \\
$\beta$             & Large scale fading        \\
$\mathbf{h}$        & Small scale fading        \\
$\mathbf{w}$        & Precoding matrix          \\
$s$                 & Transmission symbol          \\
$k$                 & The index of the $k$-th MU     \\
$n$                 & The index of the $n$-th SBS and its user \\
$(*)^m$             & The related variables of MBS DL    \\
$(*)^b$             & The related variables of backhaul DL \\
$(*)^s$             & The related variables of SBS DL      \\
\bottomrule
\end{tabular}
\end{table}

For MUs, they receive signal from the MBS while suffering the inter-tier interference from SBSs, so the received signal of MU $k$ is given as
\begin{equation}
\label{eq: received signal of MU}
y_k^m(t)=\underbrace{\sqrt{P_k^m\beta_k^m}\mathbf{h}_k^m\mathbf{w}_k^ms_k^m}_{\textmd{desired signal}}+\underbrace{\sum\limits_{n=1}^{N}\sqrt{P_n^s\beta_{nk}^s}h_{nk}^ss_n^s}_{\textmd{inter-tier\ interference}}+w_k^m.
\end{equation}
Based on the equation above, the received signal to interference and noise ratio (SINR) of MU $k$ can be written as
\begin{equation}
\label{eq: received SINR of MU}
\xi_k^m=\frac{P_k^m\beta_k^m\|\mathbf{h}_k^m\mathbf{w}_k^m\|^2}{\sum\limits_{n=1}^NP_n^s\beta_n^s{h_{kn}^s}^2+\sigma^2}.
\end{equation}

For SBSs, they receive data from the MBS and forward the data to SUs simultaneously in the same frequency band, which results in that the SBSs suffer not only the intra-tier interference from other SBSs  but also the \textit{self-interference} from themselves. As a result, the received signal of SBS $n$ can be expressed as
\begin{align}
\label{eq: received signal of small cell}
y_n^b(t)=&\underbrace{\sqrt{P_n^b\beta_n^b}\mathbf{h}_n^b\mathbf{w}_n^bs_n^b}_{\textmd{desired signal}}+\underbrace{\sum\limits_{n'=1,\neq n}^{N}\sqrt{P_{n'}^s\beta_{n'n}^b}h_{{n'}n}^bs_{n'}^s}_{\textmd{intra-tier interference}}\notag\\
&+\underbrace{\sqrt{\gamma P_n^s}s_n^s}_{\textmd{self-interference}}+w_n^b,
\end{align}
where $\sqrt{\gamma p_n^s}s_n^s$ represents the self-interference signal and the value of $\gamma$ is determined by self-interference cancellation technologies. Any self-interference cancellation technology (e.g., \cite{RWW11} and \cite{SSGBRW13}) can be applied at the SBSs, and the analysis in this paper is a general case. Thus, the received SINR of SBS $n$ can be written as
\begin{equation}
\label{eq: received SINR of small cell}
\xi_n^b=\frac{P_n^b\beta_n^b\|\mathbf{h}_n^b\mathbf{w}_n^b\|^2}{\sum\limits_{n'=1, \neq n}^NP_{n'}^s\beta_{n'n}^b{h_{n'n}^b}^2+\gamma P_n^s+\sigma^2}.
\end{equation}

For SUs, they receive signal from their associated SBSs while suffering the inter-tier interference from the MBS and the intra-tier interference from other SBSs. Mathematically, the received signal of $n$-th SBS' user can be expressed as
\begin{align}
\label{eq: received signal of SU}
&y_n^s(t)=\underbrace{\sqrt{P_n^s\beta_n^s}h_n^ss_n^s}_{\textmd{desired signal}}+\underbrace{\sum\limits_{n'=1, \neq n}^N\sqrt{P_{n'}^s\beta_{n'n}^s}h_{n'n}^ss_{n'}^s}_{\textmd{intra-tier interference}}\notag\\
&+\underbrace{\sum\limits_{k=1}^{K}\sqrt{P_k^m\beta_n^m}\mathbf{h}_n^m\mathbf{w}_k^ms_k^m+\sum\limits_{n'=1}^N\sqrt{P_{n'}^b\beta_n^m}\mathbf{h}_n^m\mathbf{w}_{n'}^bs_{n'}^b}_{\textmd{inter-tier interference}}+w_n^s.
\end{align}
Then, its received SINR can be written as
\begin{equation}
\label{eq: received SINR of SU}
\xi_n^s=\frac{P_n^s\beta_n^s{h_n^s}^2}{I_{\textmd{inter}}+\sum\limits_{n'=1, \neq n}^NP_{n'}^s\beta_{n'n}^s{h_{n'n}^s}^2+\sigma^2},
\end{equation}
where
\begin{align}
I_{\textmd{inter}}=\sum\limits_{k=1}^{K}P_k^m\beta_n^m\|\mathbf{h}_n^m\mathbf{w}_k^m\|^2+\sum\limits_{n'=1}^NP_{n'}^b\beta_n^m\|\mathbf{h}_n^m\mathbf{w}_{n'}^b\|^2.
\end{align}
Based on the Shannon capacity
\begin{equation}
\label{Shannon}
R=\log\{1+\textmd{SINR}/\omega\},
\end{equation}
the rate of access DL of MUs, the backhaul DL of SBSs and the access DL of SUs are defined by $R_k^m$, $R_n^b$, and $R_n^s$, respectively. {$\omega=2\ln(5P_e)/3$ is the SINR gap between Shannon channel capacity and a practical modulation and coding scheme achieving the BER $P_e$ \cite{C2012Energy}.} The total rate of all users can be expressed as
\begin{equation}
\mathcal{R}(\mathbf{P})=\sum\limits_{k=1}^KR_k^m+\sum\limits_{n=1}^NR_n^s,
\end{equation}
where $\mathbf{P}=[P_1^m, ..., P_K^m, P_1^b, ..., P_N^b, P_1^s, ..., P_N^s]$ is the power allocation scheme of the MBS and SBSs.

\section{Problem Formulation}
\label{Problem formulation}

In small cell networks, power allocation is an important research issue because a power allocation scheme decides the network interference (including inter-tier and intra-tier interference) level. Different from traditional wired backhaul schemes, in our proposed FD backhaul scheme, extra power is needed in the MBS to support the backhaul link. The power allocated to SBSs' backhaul link not only affects the spectrum efficiency of MUs but also decides the spectrum efficiency of SBSs. As a result, power allocation is critical in our proposed FD self-backhaul scheme of small cell networks.

The optimal power allocation policy $\mathbf{P}^*$ of the MBS and SBSs can be obtained by solving
\begin{align}
\label{optimization problem}
\max\limits_{\mathbf{P}}&\quad \mathcal{R}(\mathbf{P}) \\
\textmd{s.t.}\quad & C1: R_n^s\leq R_n^b, \forall n,\notag\\
              & C2: \sum\limits_{l=1}^{K+S}P_l^m\leq P_{\textmd{max}}^m, \notag\\
              & C3: P_n^s\leq P_{\textmd{max}}^s, \forall n,\notag\\
              & C4: R_k^m\geq R_{\textmd{min}}, \forall k,\notag\\
              & C5: R_n^s\geq R_{\textmd{min}}, \forall n.\notag
\end{align}
$C1$ specifies that the rate of SBSs' backhaul DL must be no less than that of SBS' access DL to satisfy the quality of service (QoS) of SUs. $C2$ is a transmission power constraint for the MBS in the DL and $P_{\textmd{max}}^m$ is the maximum transmission power of MBS. Similarly, $C3$ is the transmission power constraint of SBSs and $P_{\textmd{max}}^s$ is the maximum transmission power of small cell BSs. $C4$ and $C5$ specify the lowest QoS requirements of MUs and SUs, respectively.

\section{Solution to the Optimization Problem}
\label{Solution of the optimization problem}

The problem in (\ref{optimization problem}) is a non-convex problem because of the non-convexity of the objective function and the feasible set. As a result, finding a global optimum of problem (\ref{optimization problem}) is computationally expensive or even intractable. In this case, designing low-complexity algorithms to compute local optimal of problem (\ref{optimization problem}) is more meaningful in practice. In this section, we firstly reformulate the original problem (\ref{optimization problem}) into an equivalent DCP by using SCAM and an appropriate transformation of variables. Then a low-complexity CCCP-based algorithm is proposed to solve the DCP. To make the CCCP-based algorithm more effective, we propose an initial point searching algorithm to assist it. Finally, we conclude our proposed power allocation algorithm and prove its convergence.

\subsection{Re-formulation of the Original Problem}\label{section 3-1}
According to \cite{P09Information}, due to the fact that constant $1$ exits in the {Shannon Equation (\ref{Shannon})}, it is easy to lead to the objective function of (\ref{optimization problem}) being the difference of concave functions. In that way, it is difficult to solve the problem (\ref{optimization problem}). In order to avoid the difference of concave structure in objective function of (\ref{optimization problem}), we leverage the SCAM in \cite{P09Information} and extend this procedure to deal with (\ref{optimization problem}). We make use of the following lower bound
\begin{align}\label{SCALE}
\alpha\log z+\mu\leq\log(1+z)
\end{align}
that is tight with equality at a chosen value $z_0$ when the approximation constants $\alpha, \mu$ are chosen as.
\begin{equation}
\label{SCALE_tight}
\begin{aligned}
&\alpha=\frac{z_0}{1+z_0}, \\
&\mu=\log(1+z_0)-\frac{z_0}{1+z_0}\log z_0.
\end{aligned}
\end{equation}
Applying (\ref{SCALE}) and transformation $\tilde{p}=\ln p$ \cite{J02QoS, M07Wireless} to (\ref{Shannon}) results in the $R_k^m$, $R_n^b$, and $R_n^s$ relaxing to (\ref{R_k^m relax}), (\ref{R_n^b relax}) and (\ref{R_n^s relax}), respectively,
\newcounter{MYtempeqncnt}
\begin{figure*}[!t]
\normalsize
\setcounter{MYtempeqncnt}{\value{equation}}
\begin{align}
\bar{R}_k^m(\mathbf{\tilde{P}})=&\alpha_k^m\log\{\xi_k^m(\mathbf{\tilde{P}})/\omega\}+\mu_k^m\notag \\
=&\alpha_k^m\ast\left\{a\tilde{P}_k^m+\log\{\beta_k^m\|\mathbf{h}_k^m\mathbf{w}_k^m\|^2\}-\log\{\sum\limits_{n=1}^Ne^{\tilde{P}_n^s}\beta_{kn}^m{h_{kn}^m}^2+\sigma^2\}-\log\{\omega\}\right\}+\mu_k^m,\label{R_k^m relax}\\
\bar{R}_n^b(\mathbf{\tilde{P}})=&\alpha_n^b\log\{\xi_n^b(\mathbf{\tilde{P}})/\omega\}+\mu_n^b\notag \\
&=\alpha_n^b\ast\left\{a\tilde{P}_n^b+\log\{\beta_n^b\|\mathbf{h}_n^b\mathbf{w}_n^b\|^2\}-\log\{\sum\limits_{n'=1,\neq n}^Ne^{\tilde{P}_{n'}^s}\beta_{n'n}^b{h_{n'n}^b}^2+\mu e^{\tilde{P}_n^s}+\sigma^2\}-\log\{\omega\}\right\}+\mu_n^b, \label{R_n^b relax}\\
\bar{R}_n^s(\mathbf{\tilde{P}})=&\alpha_n^s\log\{\xi_n^s(\mathbf{\tilde{P}})/\omega\}+\mu_n^s \notag\\
&=\alpha_n^s\ast\left\{a\tilde{P}_n^s+\log\{\beta_n^s{h_n^s}^2\}-\log\left\{\sum\limits_{k=1}^Ke^{\tilde{P}_k^m}\beta_n^m\|\mathbf{h}_n^m\mathbf{w}_k^m\|^2+\sum\limits_{n'=1}^Ne^{\tilde{P}_{n'}^b}\beta_n^m\|\mathbf{h}_n^m\mathbf{w}_{n'}^b\|^2 \right.\right. \notag \\
& \left.\left.+\sum\limits_{n'=1,\neq n}^Ne^{\tilde{P}_{n'}^s}\beta_{n'n}^s{h_{n'n}^s}^2+\sigma^2\right\}-\log\{\omega\}\right\}+\mu_n^s, \label{R_n^s relax}
\end{align}
\setcounter{equation}{\value{MYtempeqncnt}}
\setcounter{equation}{15}
\hrulefill
\vspace*{4pt}
\end{figure*}
where all the values of $\bf\alpha$ and $\bf\mu$ are fixed, {$a=1/ln{2}$ is from the change of base of logarithms}. Then we define
\begin{equation}
\mathcal{\bar R}(\tilde{\mathbf{P}})=\sum\limits_{k=1}^K\bar{R}_k^m(\tilde{\mathbf{P}})+\sum\limits_{n=1}^N\bar{R}_n^s(\tilde{\mathbf{P}}).
\end{equation}
Furthermore, the optimization problem (\ref{optimization problem}) can be relaxed as
\begin{align}
\label{optimization DCP}
\max\limits_{\tilde{\mathbf{P}}}\quad &\bar{\mathcal{R}}(\tilde{\mathbf{P}}) \\
\textmd{s.t.}\quad &C1: \bar R_n^b(\tilde{\mathbf{P}})-\bar R_n^s(\tilde{\mathbf{P}})\triangleq \varphi_n(\tilde{\mathbf{P}})\geq 0, \forall n\notag\\
&C2: \sum\limits_{k=1}^Ke^{\tilde{\mathbf{P}}_k^m}+\sum\limits_{n=1}^Ne^{\tilde{\mathbf{P}}_n^b}\leq P_{\textmd{max}}^m,\notag\\
&C3: e^{\tilde{\mathbf{P}}_n^s}\leq P_{\textmd{max}}^s,\forall n,\notag\\
&C4: R_k^m(\tilde{\mathbf{P}})\geq R_{\textmd{min}},\forall k,\notag\\
&C5: R_n^s(\tilde{\mathbf{P}})\geq R_{\textmd{min}},\forall n.\notag
\end{align}
It is obvious that
\begin{equation}\label{relax relation}
\bar{\mathcal{R}}(\tilde{\mathbf{P}})\leq\mathcal{R}(\tilde{\mathbf{P}}),
\end{equation}
which means that problem (\ref{optimization DCP}) is maximizing the lower bound of the objective function of the problem (\ref{optimization problem}). Therefore, it is natural to iteratively tighten the bound by updating the choice of $\mathbf\alpha$ and $\mathbf\mu$ according to the new SINR values by setting
\begin{equation}\label{tight point}
[z_0]^t=[\xi(\tilde{\mathbf P})/\omega]^t \quad \forall~\xi_k^m,~\xi_n^b,~\xi_n^s,
\end{equation}
where $t$ is the iteration indicator. In each iteration,  we just need to solve problem (\ref{optimization DCP}). In this way, the original problem is simplified. In order to deal with problem (\ref{optimization DCP}), we give the following Lemma


\textbf{Lemma 1:} Given $\mathbf\alpha$ and $\mathbf\mu$, $\bar{R}_k^m(\tilde{\mathbf{P}})$, $\bar R_n^b(\tilde{\mathbf{P}})$ and $\bar R_n^s(\tilde{\mathbf{P}})$ are jointly concave with respect to the optimization variable $\tilde{\mathbf{P}}$.

\textit{Proof:} Because of the similar structure among $\bar{R}_k^m(\tilde{\mathbf{P}})$, $\bar R_n^b(\tilde{\mathbf{P}})$ and $\bar R_n^s(\tilde{\mathbf{P}})$, we take $\bar R_n^s(\tilde{\mathbf{P}})$ as an example considering that the structure of $\bar R_n^s(\tilde{\mathbf{P}})$ is most complex. We define $f(\mathbf x)=\log\sum\limits_{k=1}^Ke^{x_k}$, which decides the convexity of $\bar R_n^s(\tilde{\mathbf{P}})$. Referring to \cite{boyd2009convex}(Sec. 3.2.6), we know that $f(\mathbf x)$ is jointly convex in variables $x_k$. Due to the fact that $\bar R_n^s(\tilde{\mathbf{P}})$ is the sum of a \textit{linear} function and some \textit{concave} functions, it is jointly concave with respect to the optimization variable $\tilde{\mathbf{P}}$. The proofs of $\bar{R}_k^m(\tilde{\mathbf{P}})$ and $\bar R_n^b(\tilde{\mathbf{P}})$ are the same.

Based on Lemma 1, we know that the objective function of problem (\ref{optimization DCP}) is a strictly concave with respect to $\tilde{\mathbf{P}}$, since it is a sum of concave terms. What's more, the feasible set (defined by $\mathcal{F}$) decided by $C2, C3, C4$ and $C5$ is a convex set, since $C2$ and $C3$ are convex and $C4$ and $C5$ are concave. However, function $\varphi_n(\tilde{\mathbf{P}})$ represents the difference of two concave functions, i.e., a difference of convex decomposition \cite{an2003dc} \cite{horst1999dc}. Thus, the problem (\ref{optimization DCP}) can be seen as a DCP, which can be rewritten as
\begin{align}
\label{optimization DCP final}
\max\limits_{\tilde{\mathbf{P}}}\quad \bar{\mathcal{R}}(\tilde{\mathbf{P}})
\quad \textmd{s.t.}\quad C1, \quad \tilde{\mathbf{P}}\in \mathcal{F}.
\end{align}
\subsection{The Proposed Iterative Solution Algorithm}\label{section 3-2}
In this subsection, the CCCP, which is widely adopted for solving DCP \cite{smola2005kernel} \cite{lanckriet2009convergence}, is used to solve problem (\ref{optimization DCP final}). The main idea of the CCCP based algorithm is to iteratively approximate the original nonconvex feasible set decided by $C1$ by a convex subset and then solve the resulting convex approximation in each iteration. As the nonconvex part in problem (\ref{optimization DCP final}) stems from the fact that function $\bar R_n^s(\tilde{\mathbf{P}})$ is concave but not convex, we approximate this function in the $t$-th iteration by its first-order Taylor expansion $\hat{R}_n^s([{\tilde{\mathbf{P}}}]^t, \tilde{\mathbf{P}})$ around the current point $[{\tilde{\mathbf{P}}}]^t$. According to \cite{li2008complex}, the first-order Taylor expansion $\hat{R}_n^s$ is given by
\begin{align}
\label{Taylor equation}
\hat{R}_n^s([{\tilde{\mathbf{P}}}]^t, \tilde{\mathbf{P}})=\bar R_n^s([{\tilde{\mathbf{P}}}]^t)+\Delta \bar R_n^s([{\tilde{\mathbf{P}}}]^t)(\tilde{\mathbf{P}}-[{\tilde{\mathbf{P}}}]^t), \forall n,
\end{align}
which is an affine function about $\tilde{\mathbf{P}}$. Here, $\Delta \bar R_n^b([\tilde{\mathbf{P}}]^t)$ denotes the first-order derivative of the function $\bar R_n^b({\tilde{\mathbf{P}}}^t)$ with respect to vector $\tilde{\mathbf{P}}$.

Then, in the $t$-th iteration of the proposed CCCP based iterative algorithm, the following \textit{convex} optimization problem,
\begin{align}
\label{optimization DCP Taile}
\max\limits_{\tilde{\mathbf{P}}} \quad& \bar{\mathcal{R}}(\tilde{\mathbf{P}}) \\
\textmd{s.t.}\quad C1:\quad & \bar R_n^b(\tilde{\mathbf{P}})-\hat R_n^s(\tilde{[\mathbf{P}}]^t, \tilde{\mathbf{P}})\geq 0, \forall n, \notag\\
&\tilde{\mathbf{P}}\in\mathcal{F},\notag
\end{align}
is solved, and the solution is denoted by $[\tilde{\mathbf{P}}]^{t+1}$. This procedure is carried out iteratively until convergence or until the maximum number of allowable iterations is reached.
\begin{algorithm}[!t]
\caption{The Proposed Low-Complexity Solution}
\begin{algorithmic}[1]\label{DCP solution}
\STATE \textbf{Initialization:} Initialize maximum number of iterations $T_2^{\textmd{max}}$ and the maximum tolerance $\epsilon_2$; Initialize the algorithm with a feasible point $[\tilde{\mathbf{P}}]^0$ and set the iteration number $t$=0.
\STATE \textbf{Repeat:}
\STATE \quad Compute the affine approximation $\hat R_n^s([\tilde{\mathbf{P}}]^t, \tilde{\mathbf{P}})$ according to (\ref{Taylor equation}).
\STATE \quad Solve problem (\ref{optimization DCP Taile}), and  update $[\tilde{\mathbf{P}}]^{t+1}$.
\STATE \quad Set $t=t+1$.
\STATE \textbf{Until:} $\|\bar{\mathcal{R}}([\tilde{\mathbf{P}}]^{t+1})-\bar{\mathcal{R}}([\tilde{\mathbf{P}}]^t)\|\leq \epsilon_2$ or $t\geq T_2^{\textmd{max}}$.
\end{algorithmic}
\end{algorithm}
Since $\bar R_n^s(\tilde{\mathbf{P}})$ is concave and is approximated by its first-order Taylor expansion $\hat R_n^s([\tilde{\mathbf{P}}]^t, \tilde{\mathbf{P}})$ in problem (\ref{optimization DCP Taile}), it is obvious that
\begin{equation}\label{Taylor expansion}
\bar R_n^s(\tilde{\mathbf{P}})\leq \hat R_n^s([\tilde{\mathbf{P}}]^t, \tilde{\mathbf{P}}), \forall n,
\end{equation}
which implies that
\begin{equation}
\label{constrait approx}
\bar R_n^b(\tilde{\mathbf{P}})-\bar R_n^s(\tilde{\mathbf{P}})\geq \bar R_n^b(\tilde{\mathbf{P}})-\hat R_n^s(\tilde{\mathbf{P}}).
\end{equation}
From (\ref{constrait approx}), we know that the convex constraint $C1$ in problem (\ref{optimization DCP Taile}) can be considered as a strengthening of the original nonconvex constraint $C1$ in (\ref{optimization DCP final}). In other words, the feasible set defined in (\ref{optimization DCP Taile}) is a \textit{subset} of the true feasible set defined in (\ref{optimization DCP final}). As a result, provided that the initial point $[\tilde{\mathbf{P}}]^0$ is feasible for the DCP (\ref{optimization DCP final}), then all the iterates, $[\tilde{\mathbf{P}}]^t$ generated by iteratively solving the convex optimization problem (\ref{optimization DCP Taile}) with the affine approximation in (\ref{Taylor equation}), always belong to the true feasible set defined by $C1$ and $\mathcal{F}$ of (\ref{optimization DCP final}). We summarize the proposed low-complexity solution as Algorithm \ref{DCP solution}, where we assume that an initial feasible point $[\tilde{\mathbf{P}}]^0$ of the DCP (\ref{optimization DCP final}) is available (We will introduce how to obtain an initial feasible point in the following subsection.).

 Remark that Algorithm \ref{DCP solution} provides a low-complexity solution, in the sense that in each step a simple convex optimization problem is solved. The proposed Algorithm \ref{DCP solution} converges to a local optimum after a few iterations, as can be observed from the simulation results.
\subsection{Feasible Initial Point Searching Algorithm}\label{section 3-3}
Inspired by \cite{boyd2009convex} (Sec. 11.4),\cite{Y12Joint, B12Distributed}, we propose a feasible initial point searching algorithm, instead of an arbitrary point as in the conventional CCCP,  to obtain the feasible  $[\tilde{\mathbf{P}}]^0$ in Algorithm \ref{DCP solution}. The main advantage of the proposed new initialization method stems from the fact that, once the proposed algorithm starts with a point in the feasible set of the DCP (\ref{optimization DCP final}), all the iterates $[\tilde{\mathbf{P}}]^t$ generated by the algorithm remain within the original feasible set of the DCP (\ref{optimization DCP final}). In addition, if the CCCP is initialized with a random (infeasible) point, the CCCP may fail at the first iteration due to the infeasibility of problem. However, the task of computing a feasible point of a nonconvex optimization problem, e.g., the problem (\ref{optimization DCP final}), is NP-hard in general. This observation motivates the development of suboptimal, but low-complexity feasibility search procedures.

The proposed feasible initial point searching algorithm is based on similar iterative affine approximations of the originally nonconvex constraints as used in Algorithm \ref{DCP solution}, but with the following two modifications: a) the proposed searching algorithm starts with an arbitrary point $[\tilde{\mathbf{P}}]^0$; b) in the $t$-th iteration, instead of maximizing the spectrum efficiency of the networks as in problem (\ref{optimization DCP final}), we maximize the slack parameter $s\in\mathbb{R}$, which can be regarded as an abstract measure of the constraint violations. The feasibility problem can then be expressed as the following convex program:
\begin{align}
\label{optimization IFSP}
\max \quad& s \\
\textmd{s.t.}\quad C1:\quad & \bar R_n^b(\tilde{\mathbf{P}})-\hat R_n^s(\tilde{[\mathbf{P}}]^t, \tilde{\mathbf{P}})\geq s, \forall n, \notag\\
&\tilde{\mathbf{P}}\in\mathcal{F},\notag
\end{align}
where $\hat R_n^s(\tilde{[\mathbf{P}}]^t, \tilde{\mathbf{P}})$ is defined according to (\ref{Taylor expansion}). If the current objective value $s^{t+1}$ is zero, the algorithm stops; otherwise, the algorithm continues until convergence or until the maximum number of allowable iterations is reached. If no feasible point could be found with the proposed method, some admission control mechanisms can be adopted to reduce the number of MUs, which, however, is out of the scope of this paper. The proposed feasible initial point searching algorithm is summarized as Algorithm \ref{FISA} .
\begin{algorithm}[!t]
\caption{The Proposed Feasible Initial Point Searching Algorithm}
\begin{algorithmic}[1]\label{FISA}
\STATE \textbf{Initialization:} Initialize maximum number of iterations $T_3^{\textmd{max}}$ and the maximum tolerance $\epsilon_3$; Initialize the algorithm with a feasible point $[\tilde{\mathbf{P}}]^0$ and set the iteration number $t$=0.
\STATE \textbf{Repeat:}
\STATE \quad Compute the affine approximation $\hat R_n^s([\tilde{\mathbf{P}}]^t, \tilde{\mathbf{P}})$ according to (\ref{Taylor equation}).
\STATE \quad Solve problem (\ref{optimization IFSP}), and  update $[\tilde{\mathbf{P}}]^{t+1}$ and $[s]^{t+1}$.
\STATE \quad Set $t=t+1$.
\STATE \textbf{Until:} $[s]^{t+1}=0$ or $\|{\mathcal{R}}([\tilde{\mathbf{P}}]^{t+1})-\eta([\tilde{\mathbf{P}}]^t)\|\leq \epsilon_3$ or $t\geq T_3^{\textmd{max}}$.
\end{algorithmic}
\end{algorithm}

Note that a solution of problem (\ref{optimization IFSP}) with $s=0$ obtained is always feasible for the DCP (\ref{optimization DCP final}). Conversely, if the proposed Algorithm \ref{FISA} fails to provide a feasible point of problem (\ref{optimization DCP final}), then this does not imply that this problem is infeasible since Algorithm \ref{FISA} operates only on a subset of the original feasible set of the DCP in (\ref{optimization DCP Taile}).

The proposed Feasible Initial Point Searching Algorithm in Algorithm \ref{FISA} together with the CCCP-based Algorithm \ref{DCP solution} forms a two-step algorithm for solving the DCP in (\ref{optimization DCP final}). In the first step,  Algorithm \ref{FISA} is applied to find a feasible point of the DCP in (\ref{optimization DCP final}), instead of an random point. In the second step, the CCCP-based Algorithm \ref{SCAA} is applied, starting with the feasible point found in the first step. For convenience, the DCP solution algorithm in Algorithm \ref{DCP solution} includes the feasible initial point searching algorithm in Algorithm \ref{FISA} by default in the rest of this paper.

\subsection{Overall Algorithm and Convergence Analysis}\label{section 3-4}
From Subsection \ref{section 3-1} and Subsection \ref{section 3-2}, it can be concluded that  problem (\ref{optimization problem}) can be solved by a two-tier iteration algorithm. In the first tier, the original problem is simplified by SCAM and variables transformation, and we can approach the original problem by updating the value of $z_0$ and choosing $\alpha$ and $\mu$ according to (\ref{SCALE}) in each iteration. In this way, we just need to solve one DCP in each iteration, which makes it possible to solve problem (\ref{optimization problem}) easily. In the second tier, in order to solve the DCP, a CCCP-based iteration algorithm is proposed, in which the DCP is transformed to a convex problem by using Taylor expansion to approximate the non-convex constraint. In this way, the DCP can be solved by solving convex problems iteratively, which reduces the computational complexity significantly. The overall algorithm is summarized in Algorithm \ref{SCAA}, where the feasible initial point is obtained by a similar algorithm with Algorithm \ref{FISA} .
\begin{algorithm}[!t]
\caption{The overall Algorithm}
\begin{algorithmic}[1]\label{SCAA}
\STATE \textbf{Initialization:} Initialize maximum number of iterations $T_1^{\textmd{max}}$ and the maximum tolerance $\epsilon_1$; Initialize the algorithm with a feasible initial $\tilde{\mathbf P}$, calculate the initial $\mathbf\alpha$ and $\mathbf\mu$ according to {(\ref{SCALE_tight})}, and set the iteration number $t$=0.
\STATE \textbf{Repeat:}
\STATE \quad Solve the problem (\ref{optimization DCP}) based on Algorithm \ref{DCP solution} to obtain the current optimal $[\tilde{\mathbf{P}}]^{t+1}$.
\STATE \quad Update $\{[\mathbf\alpha]^{t+1}, [\mathbf\beta]^{t+1}\}$ according to (\ref{tight point}) and {(\ref{SCALE_tight})}.
\STATE \quad Set $t=t+1$.
\STATE \textbf{Until:} $\|\bar{\mathcal{R}}([\tilde{\mathbf{P}}]^{t+1})-\bar{\mathcal{R}}([\tilde{\mathbf{P}}]^t)\|\leq \epsilon_1$ or $t\geq T_1^{\textmd{max}}$.
\end{algorithmic}
\end{algorithm}

Next, we analyze the convergence of our proposed power allocation algorithm in FD self-backhaul small cell networks with massive MIMO. The convergence proof is carried out in two parts: 1) The convergence proof of Algorithm \ref{DCP solution} (the convergence behavior of Algorithm \ref{FISA} can be inferred accordingly); 2) the convergence proof of Algorithm \ref{SCAA}.
\subsubsection{Convergence Analysis of Algorithm \ref{DCP solution}}
We know that the point $[\tilde{\mathbf P}]^t$ is a feasible point of the convex optimization problem with concave objective function in (\ref{optimization DCP Taile}), provided that the initial point $[\tilde{\mathbf P}]^0$ is feasible for the DCP in (\ref{optimization DCP final}). As a consequence, the sequence $\{\bar{\mathcal{R}}([\tilde{\mathbf P}]^t)\}$ monotonically increases as the iteration number $t$ grows. Since the sequence $\{\bar{\mathcal{R}}([\tilde{\mathbf P}]^t)\}$ is upper-bounded by transmission power limit (C2, C3 in problem (\ref{optimization DCP})), the convergence of the sequence $\{\bar{\mathcal{R}}([\tilde{\mathbf P}]^t)\}$, and thus the convergence of Algorithm \ref{DCP solution} is guaranteed for any initial feasible point $[\tilde{\mathbf P}]^0$.

Moreover, since the objective function $\{\bar{\mathcal{R}}(\tilde{\mathbf P})\}$ of problem (\ref{optimization DCP Taile}) is strictly concave as we have proven in Subsection \ref{section 3-2}, the point $[\tilde{\mathbf P}]^{t+1}$, i.e., the solution of problem (\ref{optimization DCP Taile}), is unique \cite{boyd2009convex}. Hence, for any given initial feasible point $[\tilde{\mathbf P}]^0$, the entries of the two sequences, $\{\bar{\mathcal{R}}([\tilde{\mathbf P}]^t)\}$ and $\{[\tilde{\mathbf P}]^t\}$, have a one-to-one correspondence. As a result, the monotone convergence of the sequence $\{\bar{\mathcal{R}}([\tilde{\mathbf P}]^t)\}$ implies the convergence of the sequence $\{[\tilde{\mathbf P}]^t\}$, for any initial feasible point $\{[\tilde{\mathbf P}]^0\}$. Let $\tilde{\mathbf P}^*([\tilde{\mathbf P}]^0)$ denote the limit point of the sequence $\{[\tilde{\mathbf P}]^t\}$ with a feasible initialization $\{[\tilde{\mathbf P}]^0\}$ when the iteration number $t$ goes to infinity, i.e., given the initial feasible point $\{[\tilde{\mathbf P}]^0\}$, we have
\begin{equation}\label{eq: limit point defination}
\tilde{\mathbf P}^*([\tilde{\mathbf P}]^0)\triangleq\lim\limits_{t\rightarrow\infty}[\tilde{\mathbf P}]^t
\end{equation}
In general, the limit point $\tilde{\mathbf P}^*([\tilde{\mathbf P}]^0)$ depends on the choice of the initial feasible point $\{[\tilde{\mathbf P}]^0\}$. For notational simplicity, we write the limit point as $\tilde{\mathbf P}^*$. Regarding the limit point $\tilde{\mathbf P}^*$, we have the following lemma.

\textbf{Lemma 2:} The limit point $\tilde{\mathbf P}^*$ of $\{[\tilde{\mathbf P}]^t\}$ is the solution of the following convex optimization problem:
\begin{align}\label{eq: limit point solution}
\max\limits_{\tilde{\mathbf P}}\quad &\bar{\mathcal{R}}(\tilde{\mathbf P}) \\
\textmd{s.t.}\quad &C1: \bar R_n^b(\tilde{\mathbf P})-\hat{R}_n^s(\tilde{\mathbf P}^*, \tilde{\mathbf P})\geq0, \ \forall n,\notag \\
&\tilde{\mathbf P}\in \mathcal{F},\notag
\end{align}
where the affine function $\hat{R}_n^s(\tilde{\mathbf P}^*, \tilde{\mathbf P})$ is obtained by replacing $\{[\tilde{\mathbf P}]^t\}$ with $\tilde{\mathbf P}^*$ in (\ref{constrait approx}). Moreover, the limit point $\tilde{\mathbf P}^*$ satisfies all the constraints C1 in (\ref{eq: limit point solution}) with equalities, i.e.,
\begin{equation}\label{eq: limit point C1 tight}
\bar R_n^b(\tilde{\mathbf P}^*)-\hat{R}_n^s(\tilde{\mathbf P}^*, \tilde{\mathbf P}^*)=\bar R_n^b(\tilde{\mathbf P}^*)-\bar{R}_n^s(\tilde{\mathbf P}^*)=0.
\end{equation}

\textit{Proof:} By definition (\ref{eq: limit point defination}), the point $\tilde{\mathbf P}^*$ is the limit point of the sequence $[\tilde{\mathbf P}]^t$, hence the point $\tilde{\mathbf P}^*$ is a feasible point for the convex optimization problem (\ref{eq: limit point solution}) and no strictly better solution exists. What's more, as we have proven in \ref{section 3-2}, the objective function $\bar{\mathcal{R}}(\tilde{\mathbf P})$ in (\ref{eq: limit point solution}) is strictly concave in the variable $\tilde{\mathbf P}$, so the solution of the problem (\ref{eq: limit point solution}) is unique \cite{boyd2009convex} (Sec. 4.2), which means that the limit point $\tilde{\mathbf P}^*$ is the solution of problem (\ref{eq: limit point solution}). We prove the second part of the Lemma by contradiction. Assuming that the constraint C1 of the $n$-th SBS is not active, i.e., $\bar R_n^b(\tilde{\mathbf P})-\hat{R}_n^s(\tilde{\mathbf P}^*, \tilde{\mathbf P})>0$, we can scale down the variable $\tilde{P}_n^b$ to make the constraint active without violating the other constraints, which makes it possible that the MUs or other SBSs' backhaul link can be allocated more power or lower interference level. Then the objective function will increase possibly, which contradicts the optimality of the point $\tilde{\mathbf P}^*$. Hence, it can be concluded that all constraints in (\ref{eq: limit point solution}) C1 are active at the point $\tilde{\mathbf P}^*$.

According to the Lemma 2, no matter how to choose the initial point $[\tilde{\mathbf P}]^0$, only if it is feasible, the final convergence point can obtained by solving the problem (\ref{eq: limit point solution}). In other words, the limit point $\tilde{\mathbf P}^*$ is a stationary point of the DCP (\ref{optimization DCP}) \cite{W2004operations}. Therefore, we have the conclusion that our proposed Algorithm \ref{DCP solution} not only converges but also converges to a stationary point.

\subsubsection{Convergence Analysis of Algorithm \ref{SCAA}}
We start the convergence analysis of Algorithm \ref{SCAA} with the following proposition.

\textbf{Proposition 1:} Algorithm \ref{SCAA} monotonically improves the value of the objective function at each iteration and converges. The solution obtained upon convergence satisfies the necessary optimality conditions.

\textit{Proof:} Let $[\tilde{\mathbf P}]^t$, $[\bm\alpha]^t$, and $[\bm\mu]^t$ be the optimized values of the $t$-th iteration. Then, we have
\begin{align}
\cdots\leq\bar{\mathcal{R}}([\tilde{\mathbf P}]^t)\overset{(a)}{=}\bar{\mathcal{R}}_{[\bm\alpha]^t, [\bm\mu]^t}([\tilde{\mathbf P}]^t)\overset{(b)}{\leq}\bar{\mathcal{R}}_{[\bm\alpha]^t, [\bm\mu]^t}([\tilde{\mathbf P}]^{t+1})\notag\\
\overset{(c)}{\leq}\bar{\mathcal{R}}([\tilde{\mathbf P}]^{t+1})\overset{(a)}{=}\bar{\mathcal{R}}_{[\bm\alpha]^{t+1}, [\bm\mu]^{t+1}}([\tilde{\mathbf P}]^{t+1})\leq\cdots,
\end{align}
where the quality (a) is due to the fact that the relaxations of (\ref{R_k^m relax})(\ref{R_n^b relax})(\ref{R_n^s relax}) are tight at the current SINR values according to (\ref{SCALE}) and (\ref{tight point}), the inequality (b) is due to the fact that the maximization (\ref{optimization DCP}) is strictly concave, the inequality (c) follows from (\ref{relax relation}). For a finite set of transmit sum powers and channel gains, since the optimal spectrum efficiency is bounded above, the procedure must converge.

As a conclusion, the proposed power algorithm in FD small cell networks with massive MIMO converge well.
We also evaluate its convergence performance by simulations in the next section.
\section{Simulation Results and Discussions}
\label{Simulation}
\begin{table}[!t]
\caption{The simulation parameters}
\label{simulation parameters}
\centering
\begin{tabular}{ll}
\toprule
Simulation parameters & Value\\
\midrule
The number of the MBS' antennas $M$ & 128 \\
Path loss exponent & $-3$ \\
Power spectral density of noise & $-174\textmd{dBm}/Hz$ \\
Circuit power consumption $P_C$ & $160\textmd{mW}$ \\
Power amplifier efficiency $1/\rho$ & $38\%$ \\
The power consuming weight $w$ & $10$ \\
The maximum transmission power of the MBS & $46\textmd{dBm}$ \\
The maximum transmission power of SBSs & $20\textmd{dBm}$ \\
The QoS requirement of MUs & 2 bit/s/Hz \\
\bottomrule
\end{tabular}
\end{table}
In this section, the effectiveness of our proposed FD self-backhaul scheme of small cell networks with massive MIMO will be demonstrated by Monte Carlo simulations, where the simulation results are averaged over 1000 droppings. In the simulations, we consider a $0.5\textmd{Km}\times0.5\textmd{Km}$ square area covered by one MBS located in the center and some SBSs (pico base stations are adopted) that are randomly deployed.  The simulation parameters used are listed in Table \ref{simulation parameters}, which are similar to those in  \cite{L14O}.
\subsection{Performance Comparison with Existing Schemes}
We evaluate the performance of the proposed backhaul scheme by comparing the following schemes: (a) a traditional wired backhaul scheme without massive MIMO \cite{Y14W}, where the MBS schedules MUs in time domain; (b) an existing FD backhaul scheme without massive MIMO \cite{SSGBRW13}, where the MBS allocates the orthogonal frequency band to MUs and SBSs' backhaul link; (c) a variation of our proposed scheme with massive MIMO and half duplex (HD), where MBS can serve MUs and SBSs' backhaul link simultaneously, but each SBS receives and transmits date in different time slots. Each scheme has the similar system configurations as described above. In this subsection, we assume the self-interference is canceled incompletely ($\gamma=10^{-5}$) \cite{SSGBRW13}.
\begin{figure}[!t]
\centering
\includegraphics[width=0.5\textwidth]{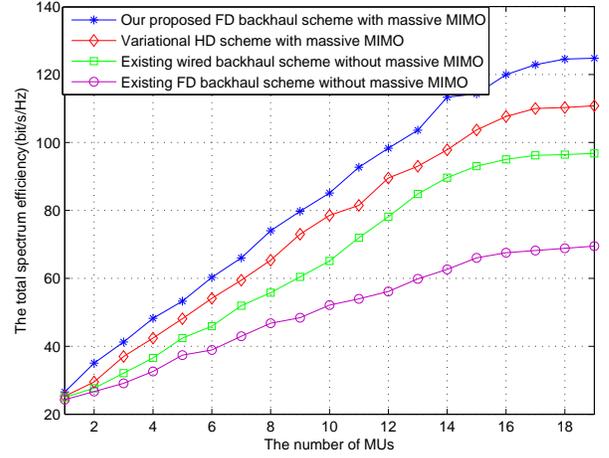}
\caption{The spectrum efficiency performance of different backhaul schemes with different numbers of MUs ($N=4$).}
\label{fig: Total-SE-num_MU}
\end{figure}

\begin{figure}[!t]
\centering
\includegraphics[width=0.5\textwidth]{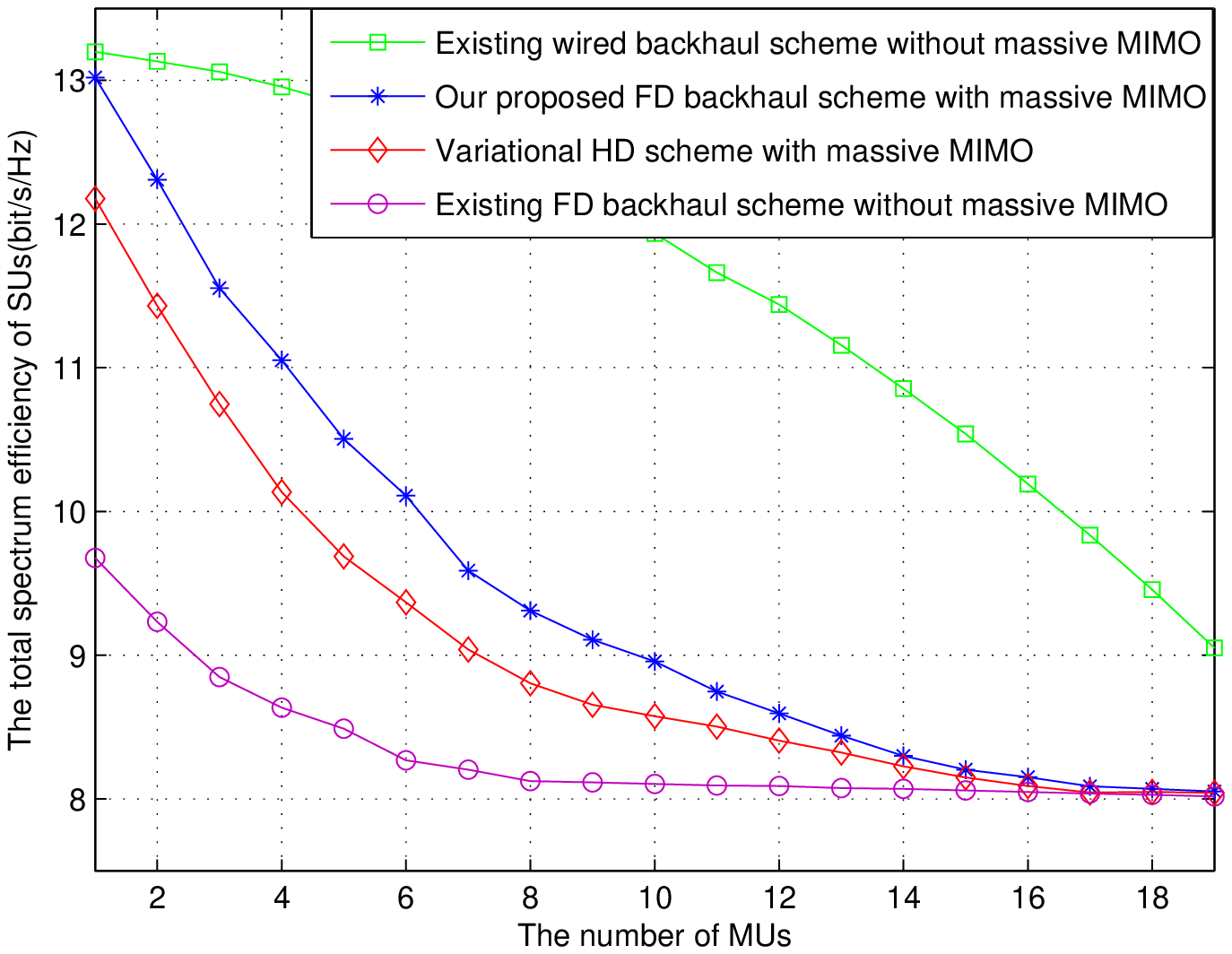}
\caption{The total spectrum efficiency of SUs of different backhaul schemes with different numbers of MUs ($N=4$).}
\label{fig: SBS-SE-num_MU}
\end{figure}

\begin{figure}[!t]
\centering
\includegraphics[width=0.5\textwidth]{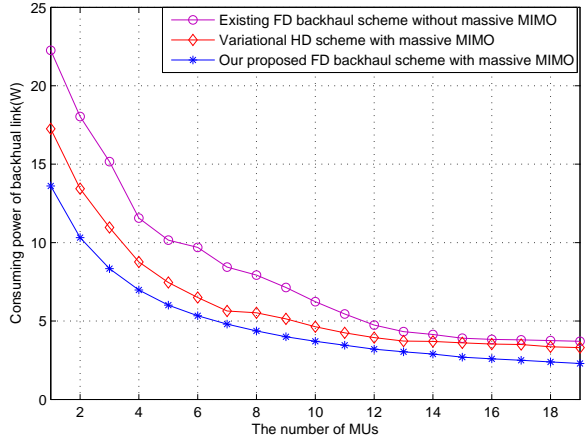}
\caption{The consuming power of backhaul link of different backhaul schemes with different numbers of MUs ($N=4$).}
\label{fig: Backhaul-P-num_MU}
\end{figure}

\subsubsection{The effect of the number of MUs}
In Figs. \ref{fig: Total-SE-num_MU}, \ref{fig: SBS-SE-num_MU}, \ref{fig: Backhaul-P-num_MU}, we compare the performance of different backhaul schemes with different numbers of MUs. As shown in Fig. \ref{fig: Total-SE-num_MU}, the spectrum efficiency performance of our proposed backhaul scheme outperforms the HD self-backhaul scheme with massive MIMO, the wired backhaul scheme without massive MIMO, and the FD self-backhaul scheme without massive MIMO. This is because our proposed backhaul scheme can take the advantage of both massive MIMO and FD backhaul. By massive MIMO technology, the backhaul DL of SBSs could be completed in the same frequency band at the same time with the MUs, rather than on the orthorhombic frequency band like in the schemes without massive MIMO, which can improve SE. By FD technology, the SBSs can receive backhaul data and transmit DL data simultaneously instead of in a time division pattern like in the HD backhaul scheme, which improves spectrum efficiency further. In addition, our proposed power allocation algorithm mitigates the inter-tier and intra-tier interference, which also has contribution to the high spectrum efficiency of our proposed backhaul scheme. What's more, it can be observed from Fig. \ref{fig: Total-SE-num_MU} that the increasing rate of spectrum efficiency of all backhaul schemes becomes lower when the number of MUs reaches to some values. Actually, these inflection points mean the maximum number of serving MUs of different backhaul schemes. When the number of MUs exceeds the maximum number of serving MUs, the spectrum efficiency will continue to increase because of the multi-user diversity gain. Note that the maximum number of serving MUs is the highest in our proposed FD backhaul scheme with massive MIMO.

Fig. \ref{fig: SBS-SE-num_MU} and Fig. \ref{fig: Backhaul-P-num_MU} show the change of the total spectrum efficiency of all SUs and the consuming power of backhaul link with the increase of the number of MUs, respectively. As shown in these two figures, for the wireless backhaul scheme, both the total spectrum efficiency of all SUs and the consuming power of backhaul link decrease when the number of MUs increases. The reason is that more MUs with better channel quality will encroach the power that should be allocated SBSs' backhaul link, and  the spectrum efficiency of SUs will decrease. However, because of the limit of the lowest QoS requirement, the spectrum efficiency of SUs will not be zero. Comparing with FD backhaul scheme, the HD backhaul scheme only has half of time to transmit data to SUs and another half of time to receive data from MBS, so the spectrum efficiency of SUs of HD backhaul scheme will be low and the backhaul link will consume more power to reach the level of SE. In the non-massive MIMO backhaul scheme, besides power, the MBS needs to allocate orthogonal frequency band to SBSs' backhaul link, and then the available frequency band will reduce with the increase of the number of MUs, which results in the poor spectrum efficiency performance of SUs and consuming more backhaul power to satisfy the QoS. For the wired backhaul scheme, when the number of MUs increases, more MUs will appear near SBSs and they will aggravate the inter-tier interference to SUs, so the spectrum efficiency of SUs of wired backhaul scheme will decrease but the decreasing rate will be slow  comparing with the wireless backhaul scheme. From Fig. \ref{fig: SBS-SE-num_MU} and Fig. \ref{fig: Total-SE-num_MU}, it can be observed that the spectrum efficiency of SUs of wired backhaul scheme will exceed other wireless backhaul scheme when the number of MUs is equal to about $6$ but the total spectrum efficiency of the wireless backhaul scheme with massive MIMO  is still higher than that of wired backhaul. This is because the MUs' spectrum efficiency gain obtained from massive MIMO is higher than the SUs' spectrum efficiency loss resulting from the less backhaul power when the number of MUs increases.

\begin{figure}[!t]
\centering
\includegraphics[width=0.5\textwidth]{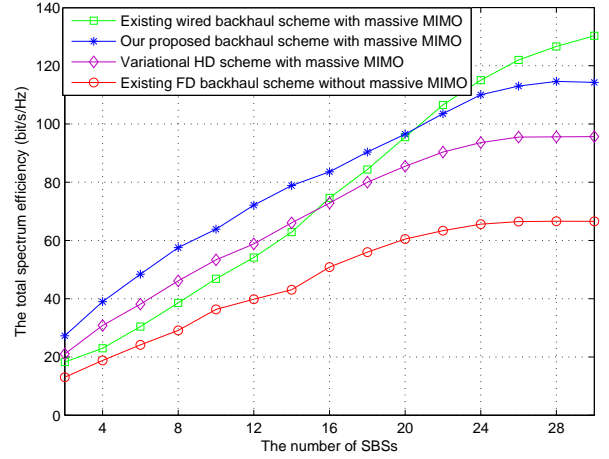}
\caption{The total spectrum efficiency of different backhaul schemes with different numbers of SBSs ($K=4$).}
\label{fig: SE-num_SBS}
\end{figure}

\begin{figure}[!t]
\centering
\includegraphics[width=0.5\textwidth]{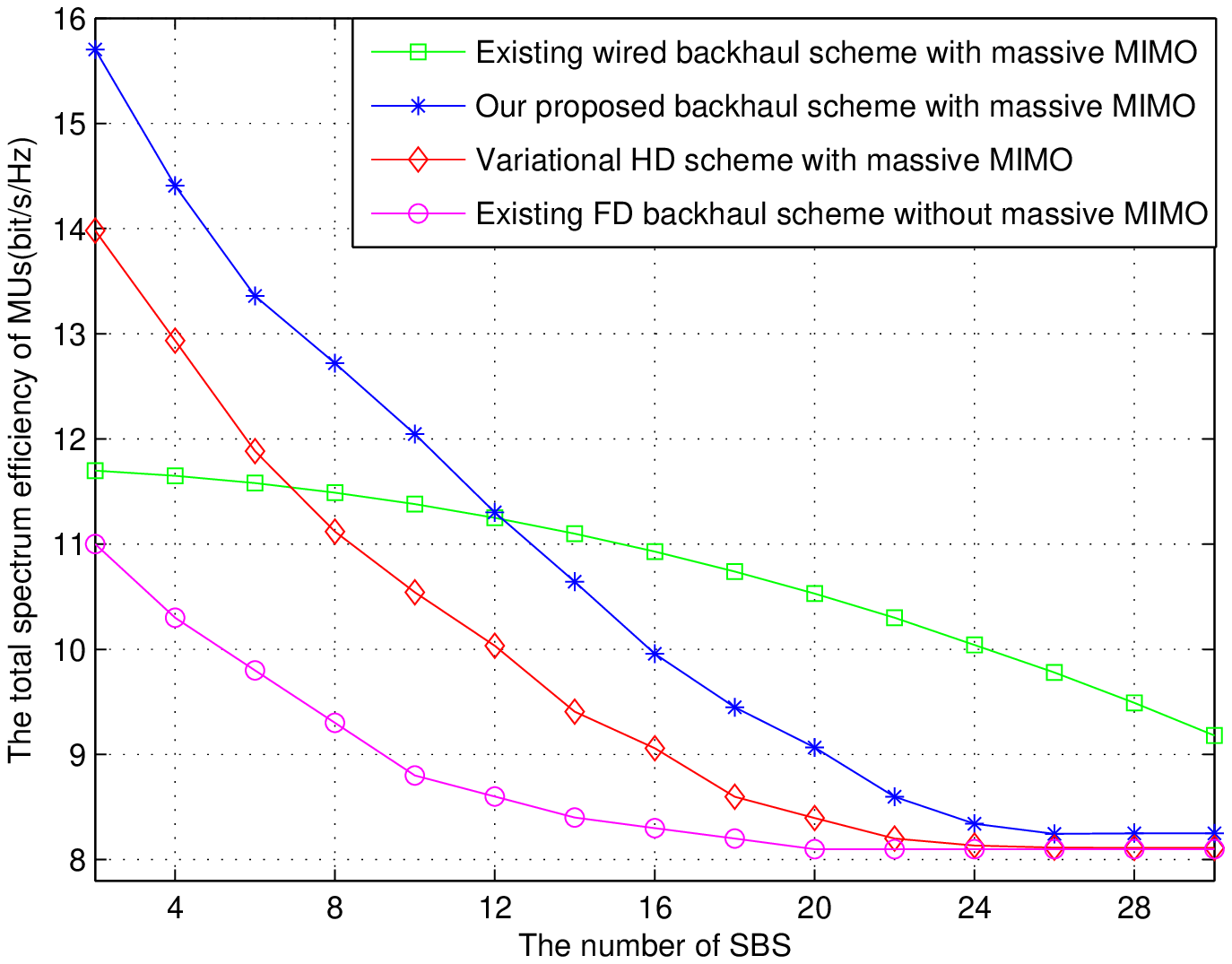}
\caption{The total spectrum efficiency of MUs of different backhaul schemes with different numbers of SBSs ($K=4$).}
\label{fig: MBS_SE-num_SBS}
\end{figure}

\begin{figure}[!t]
\centering
\includegraphics[width=0.5\textwidth]{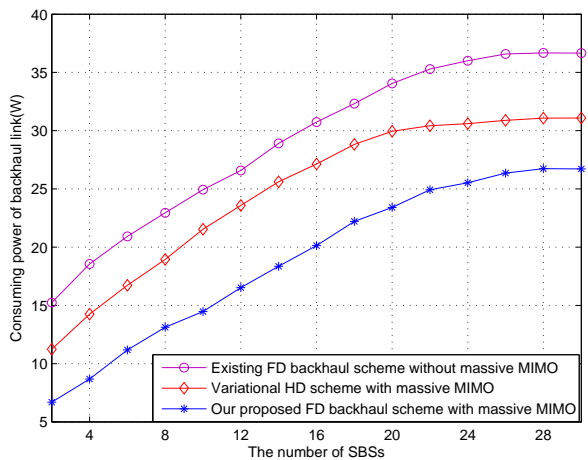}
\caption{The consuming power of backhaul link of different backhaul schemes with different numbers of SBSs ($K=4$).}
\label{fig: Backhaul_P-num_SBS}
\end{figure}
\subsubsection{The effect of the number of SBSs }
In Figs. \ref{fig: SE-num_SBS}, \ref{fig: MBS_SE-num_SBS} and \ref{fig: Backhaul_P-num_SBS}, we compare the total SE, spectrum efficiency of MUs and consuming power of backhaul link of different backhaul schemes with the increase of the number of SBSs, respectively. As shown in Fig. \ref{fig: SE-num_SBS}, the spectrum efficiency performance of all backhaul schemes firstly increases but the increasing rate decreases with the number of SBSs growing. The common reason is that the interference (inter-tier and intra-tier) level is relatively low when there are less SBSs and the interference will be serious when there are  more SBSs. For wireless backhaul schemes, another main reason is that the backhaul power of each SBS will reduce with the increase of the number of SBSs due to the fact that the power of MBS is limited. This can be verified in Fig. \ref{fig: Backhaul_P-num_SBS}, where consuming power of backhaul link will not grow any more when the number of SBSs is relatively large. At the same time, from Figs. \ref{fig: SE-num_SBS}, \ref{fig: MBS_SE-num_SBS} and \ref{fig: Backhaul_P-num_SBS}, we can find that the performance of our proposed self-backhaul scheme outperforms the HD self-backhaul scheme with massive MIMO and the FD self-backhaul scheme without massive MIMO. The reason is the same with that in Figs. \ref{fig: Total-SE-num_MU}, \ref{fig: SBS-SE-num_MU}, \ref{fig: Backhaul-P-num_MU}. However, different form Fig. \ref{fig: Total-SE-num_MU} and Fig. \ref{fig: SBS-SE-num_MU}, the spectrum efficiency performance of our proposed FD self-backhaul scheme is better than that of the wired backhaul scheme without massive MIMO when the number of SBSs is less, but the spectrum efficiency performance of the wired backhaul scheme without massive MIMO will be better than our scheme when the number of SBSs increases to a relatively higher value. This is because the backhaul link needs to consume radio power in our scheme but it does not consume radio power in the wired backhaul scheme. When the number of SBSs is less, the backhaul consumes less power because of low interference level and lower QoS requirement of SUs. With the increase of the number of SBSs, the interference level and the QoS requirement of SUs will grow, which leads to consuming more backhaul power to satisfy the QoS requirement of SUs. This is also the reason why the spectrum efficiency of MUs keeps falling until the lowest QoS requirement is reached in Fig. \ref{fig: SBS-SE-num_MU}. When the spectrum efficiency gain of SUs obtained by improving the number of SUs can not cover the spectrum efficiency cost of MUs because of the less available power, the spectrum efficiency performance of our proposed scheme will be worse than the wired backhaul scheme. Fortunately, our scheme is economic, and the wired backhaul is expensive. So a tradeoff exits between the network cost and network performance.
\subsection{The Effect of Self-interference Cancellation Performance}
\begin{figure}[!t]
\centering
\includegraphics[width=0.5\textwidth]{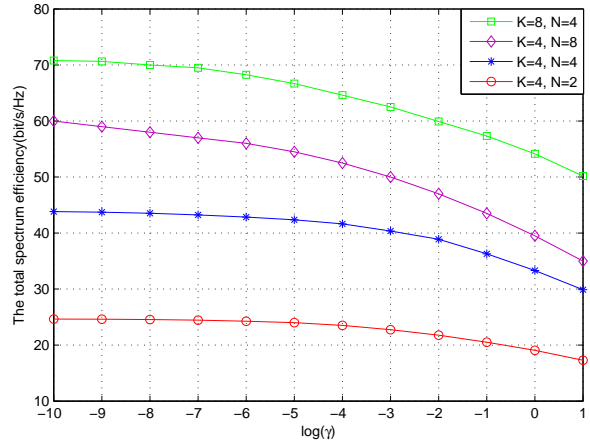}
\caption{The spectrum efficiency performance of different SI cancellation performance with different numbers of MUs and SBSs.}
\label{fig: SE-SI}
\end{figure}
In Fig. \ref{fig: SE-SI}, we study the effect of self-interference cancellation performance on our proposed FD self-backhaul scheme. As shown in Fig. \ref{fig: SE-SI}, the total spectrum efficiency will decrease with the increase of the value of $\gamma$ and the decreasing rate becomes fast. This is because a larger value of $\gamma$ means more serious self-interference, which results in the increase of consuming power of backhaul link. Consequently, the available power for MUs will be less, and the spectrum efficiency of MUs will decrease. At the same time, the MBS also will allocate less power to SBSs' backhaul link since the MBS will think the channel quality of those links is bad, which will influence the spectrum efficiency of SUs. As a result, the total spectrum efficiency performance will be bad when the self-interference is not cancelled perfectly. What's more, it can be observed from Fig. \ref{fig: SE-SI} that the total spectrum efficiency will be more sensitive to $\gamma$ when there are more SBSs.
\subsection{Convergence Illustration of the Proposed Algorithms}
\begin{figure}[!t]
\centering
\includegraphics[width=0.5\textwidth]{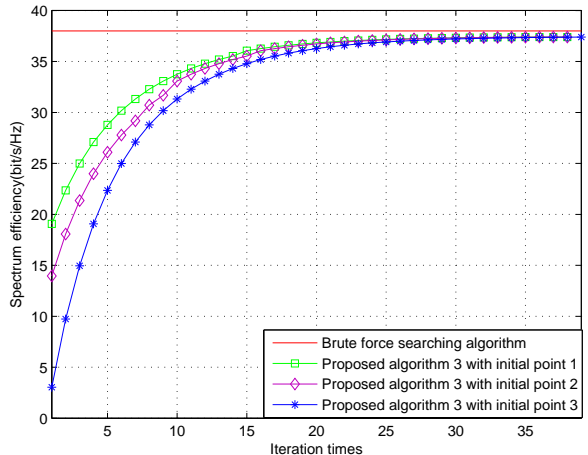}
\caption{The convergence of proposed overall algorithm with different initial point ($N=4$, $K=4$).}
\label{fig: convergence_SCAM}
\end{figure}
Fig. \ref{fig: convergence_SCAM} shows the convergence behavior of the proposed power allocation algorithm in Algorithm \ref{SCAA} with different initial feasible points, as well as the optimal solution obtained by the brute force searching (BFS) algorithm, for the FD self-backhaul small cell networks including 4 MUs and 4 SBSs. As shown in this figure, we can observe the good convergence performance and robustness to initial points. No matter where the initial point is, our algorithm will converge. What's more, the gap between Algorithm \ref{SCAA} and the BFS algorithm is narrow after sufficient iterations, although our solution is not a global optimal solution, which means that the proposed Algorithm \ref{SCAA} is effective. Furthermore, it can be observed from this figure that a significant decrease of gap between Algorithm \ref{SCAA} and BFS algorithm can be found from the first iteration to the 10-th iteration. After the 10-th iteration, the gain of more iterations is still increasing but with less rate. Thus, a tradeoff exits between the acceptable utility value and iteration steps.

\begin{figure}[!t]
\centering
\includegraphics[width=0.5\textwidth]{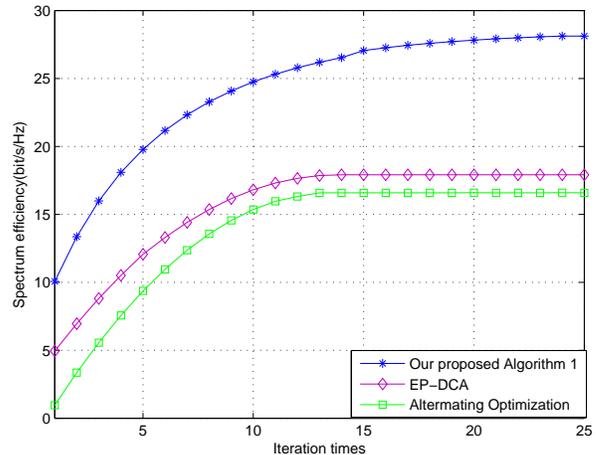}
\caption{The convergence of proposed CCCP-based algorithm ($N=4$, $K=4$).}
\label{fig: convergence_CCCP}
\end{figure}

In Fig. \ref{fig: convergence_CCCP}, the convergence behavior of the proposed Algorithm \ref{DCP solution} and two reference algorithms, i.e., the alternating optimization scheme of \cite{Ying2010Joint} and the EP-DCA of \cite{an2003dc}, are studied for the proposed FD self-backhaul small cell networks with 4 MUs and 4 SBSs. As shown in this figure, the proposed Algorithm \ref{DCP solution} converges after approximately 20 iterations for any considered initial feasible point but the alternating optimization scheme of \cite{Ying2010Joint} and the EP-DCA of \cite{an2003dc} converge after approximately 13 iterations, which implies that the convergence performance of the proposed Algorithm \ref{DCP solution} is worse than that of the two reference algorithms. However, it is obvious that the optimization performance of our proposed Algorithm \ref{DCP solution} outperforms the two reference algorithms, which is the reason why the performance of Algorithm \ref{SCAA} is close to the BFS algorithm.

\section{Conclusion and Future Work}\label{Sec 6: conclusion}
\label{Conclusion}
In this paper, we proposed a self-backhaul scheme for small cell networks with massive MIMO and FD, which enables the access of users and backhaul of SBSs simultaneously in the same frequency band. Furthermore, in order to mitigate the inter-tier and the intra-tier interference, we formulated the power allocation problem as an optimization problem, in which spectrum efficiency was taken as the optimization objective. Considering the high computation complexity for solving the non-convex optimization problem, we introduced the SCAM and an appropriate transformation of variables to transform equivalently the original problem into a DCP, which can be efficiently solved with local optimality using a CCCP-based algorithm. Simulation results showed that the proposed self-backhaul scheme by jointly using massive MIMO and FD technology is able to take the advantages of both massive MIMO and in-band FD communications. In addition, simulation results also demonstrated the effectiveness and good convergence performance of our proposed CCCP-based power allocation algorithm. Future work is in progress to consider the optimization of massive MIMO precoding matrix in our proposed scheme.

\bibliography{r}

\end{document}